\documentclass{article}
\usepackage{graphicx}  
\usepackage{amsmath}   
\usepackage[compress]{cite}
\usepackage{amssymb}   
\usepackage{bm} 
\usepackage{dcolumn}
\usepackage{color}
\usepackage{mathrsfs}
\usepackage{amsfonts}
\usepackage{varioref}
\usepackage{float}
\usepackage[caption=false]{subfig}
\RequirePackage[colorlinks,citecolor=blue,urlcolor=magenta,linkcolor=blue]{hyperref}
\input epsf
\usepackage{tablefootnote}
\def\gr{general relativity}

\labelformat{section}{Section #1} 
\labelformat{subsection}{Section #1} 
\labelformat{subsubsection}{Section #1}
\labelformat{subsubsubsection}{Section #1}
\labelformat{equation}{Eq.~(#1)} 
\labelformat{figure}{Fig.~#1} 
\labelformat{subfigure}{Fig.~\thefigure#1} 
\labelformat{table}{Table~#1} 
\labelformat{appendix}{Appendix #1}
\title{\bf Deciphering signatures of Bardeen black holes from the observed quasi-periodic oscillations}

\author{Indrani Banerjee\footnote{banerjeein@nitrkl.ac.in}~$^{1}$\\
{\small{$^{1}$Department of Physics and Astronomy, National Institute of Technology, Rourkela-769008, India}}}
\date{ }  
\begin{document}
  
\maketitle
\begin{abstract}
Quasi-periodic oscillations (QPOs) observed in the power spectrum of black holes are unique observational probes to the background spacetime since they can be directly related to the timescales associated with the motion of matter orbiting in the vicinity of the black hole horizon. In this regard, the high frequency QPOs (HFQPOs) are particularly interesting as they occur in commensurable pairs, the most common ratio being the 3:2 twin peak QPOs. The theoretical models which aim to explain these QPOs express the observed frequencies in terms of the epicyclic motion of test particles in a given background spacetime. In this work, we study the signatures of Bardeen spacetime from the observed QPOs in the black hole power spectrum. Bardeen black holes are rotating, regular black holes with a magnetic monopole charge. Such regular backgrounds are theoretically interesting as they can potentially evade the curvature singularity, otherwise unavoidable in general relativistic black holes. We perform a $\chi^2$ analysis by comparing the available observations of the quasi-periodic oscillations from black hole sources with the relevant theoretical models and note that the Kerr black holes in \gr\ are observationally more favored compared to black holes with a monopole charge. Our analysis reveals that black holes with very high monopole charges are \emph{disfavored} from QPO related observations. 
\end{abstract}
\section{Introduction}\label{QPO_Intro}

Despite the astounding success of \gr\ in explaining a plethora of observations, e.g. perihelion precession of mercury, bending of light, gravitational redshift \cite{Will:2005yc,Will:1993ns,Will:2005va,Berti:2015itd}, black hole image \cite{Fish:2016jil,Akiyama:2019cqa,Akiyama:2019brx,Akiyama:2019sww,Akiyama:2019bqs,Akiyama:2019fyp,Akiyama:2019eap} and gravitational waves \cite{Abbott:2017vtc,TheLIGOScientific:2016pea,Abbott:2016nmj,TheLIGOScientific:2016src,Abbott:2016blz}, there remains certain unresolved issues in both observational and theoretical fronts. In the observational end it cannot address adequately the dark sector \cite{Milgrom:1983pn,Milgrom:2003ui,Bekenstein:1984tv,Clifton:2011jh,Perlmutter:1998np,Riess:1998cb} while in the theoretical front \gr\ still cannot explain the big bang and black hole singularities \cite{Penrose:1964wq,Hawking:1976ra,Wald,Christodoulou:1991yfa}. In this regard, study of regular black holes is important as it can provide a possible resolution to the black hole singularity problem. Regular black holes arise in several alternative gravity theories, one such scenario being study of Einstein gravity in presence of non-linear electrodynamics. In this regard we study Bardeen black holes \cite{Bardeen}, which correspond to black hole with a magnetic monopole charge. 

In order to investigate observational signatures of Bardeen metric we consider quasi-periodic oscillations (QPOs) observed in the power spectrum of black holes \cite{2006csxs.book.....L,vanderKlis:2000ca}. In particular, the high frequency quasi-peiodic oscillations (HFQPOs) are interesting as these can be used to decipher the nature of the background spacetime. 
This is because, the observed frequency range of QPOs varies from mHz (for supermassive black holes) to hundreds of Hz (for stellar-mass black holes) which in turn is associated with dynamical timescales ($\sim 0.1–1~ \rm ms$) of motion of accreting matter very close to the black hole event horizon. One can heuristically derive such dynamical timescales $t_{\rm d}\sim \sqrt{(r^{3}/GM)}$ from characteristic velocities $v\sim \sqrt{(GM/r)}$ of motion of accreting matter near the central object. Considering, $r\sim 100~\textrm{km}$ for a $10~M_\odot$ black hole, the dynamical timescale turns out to be $t_{\rm d}\sim 1~\textrm{ms}$. 
This estimation shows that the dynamical timescales are in the millisecond range, predicted back in the 1970s \cite{1971SvA....15..377S,1973SvA....16..941S} and received observational confirmation with the launch of NASA's Rossi X-Ray Timing Explorer satellite \cite{2006csxs.book.....L}.

Several theoretical models exist in the literature which aims to explain these observed HFQPOs \cite{Stella:1997tc,PhysRevLett8217,Stella_1999,Cadez:2008iv,Kostic:2009hp,Germana:2009ce,Kluzniak:2002bb,Abramowicz:2003xy,Rebusco:2004ba,Nowak:1996hg,Torok:2010rk,Torok:2011qy,Kotrlova:2020pqy,1980PASJ...32..377K,Perez:1996ti,Silbergleit:2000ck,Dexter:2013sxa,Rezzolla:2003zy,Rezzolla:2003zx}. These requires one to study the epicyclic motion of test particles around these compact objects. The theoretically predicted QPO frequencies in all of these models turn out to be linear combinations of the angular frequency and the epicyclic frequencies. In next section we discuss the Bardeen black hole solution, i.e. Einstein gravity in non-linear electrodynamics. In \ref{S3} we derive the epicyclic frequencies in a general stationary axisymmetric spacetime. The existing QPO models are briefly discussed in \ref{S4}. In \ref{S5} we compare the theoretical QPO frequencies obtained from each of these models with the available observations. We perform a $\chi^2$ analysis to derive the observationally favored magnetic monopole charge parameter. We conclude with a summary of our results with some scope for future work in \ref{S6}. It may be important to note that we work with mostly positive metric convention and consider geometrized units i.e., $G=1=c$.

\section{Basics of Bardeen rotating black holes}\label{S2}
The line element corresponding to Bardeen black hole \cite{Bardeen} in Boyer-Lindquist coordinates is given by,
\begin{align}\label{metric_bardeen}
ds^{2} &=-\bigg{(} 1 - \frac{2\tilde{m}(r)r}{\tilde{\Sigma}}\bigg{)}dt^{2} - \frac{4a\tilde{m}(r)r}{\tilde{\Sigma}}\sin^{2}\theta dt d\phi + \frac{\tilde{\Sigma}}{\Delta}dr^{2} \nonumber\\
&+\tilde{\Sigma} d\theta^{2} + \bigg{(} r^{2} + a^{2} + \frac{2\tilde{m}(r)ra^{2}}{\tilde{\Sigma}}\sin^{2}\theta\bigg{)}\sin^{2}\theta d\phi^{2}
\end{align}
where, 
\begin{align}\label{metricparams_bardeen}
\tilde{\Sigma} = r^{2} + a^{2}\cos^{2}\theta ~ {,} ~ \Delta = r^{2} + a^{2} - 2\tilde{m}(r)r
\end{align}
and $\tilde{m}(r)r$ is the mass function such that $lim_{r\rightarrow\infty}\tilde{m}(r) = \tilde{\mathcal{M}}$ and $a$ is the Kerr parameter. The mass function corresponding to the Bardeen rotating black hole is given by,
\begin{align}\label{massfn}
\tilde{m}(r) = \tilde{\mathcal{M}}\bigg{[} \frac{r^2}{r^2 + \tilde{g}^{2}}\bigg{]}^{3/2}
\end{align}
where $\tilde{g}$ is the magnetic monopole charge. 
Due to the presence of the magnetic monopole charge Bardeen black hole has no curvature singularity, i.e., spacetime is regular everywhere and it satisfies the weak energy condition.  

The black hole solution corresponding to \ref{metric_bardeen} arises in nonlinear electrodynamics\cite{Bardeen}, the associated action being given by,
\begin{align}\label{action_bardeen}
\mathcal{S} = \int ~ d^4x \bigg{[} \frac{\mathcal{R}}{16\pi} - \frac{\mathcal{W}(f)}{4\pi}\bigg{]}
\end{align}
In \ref{action_bardeen} $\mathcal{R}$ is the Ricci scalar, $\mathcal{W}(f)$ is a function of $f = \frac{1}{4}\mathcal{F}_{\mu\nu}\mathcal{F}^{\mu\nu}$ where $\mathcal{F}_{\mu\nu} = 2\nabla_{[\mu}A_{\nu]}$ as the electromagnetic field strength. Varying the above action with respect to the metric we obtain the following gravitational field equations:
\begin{align}
&\mathcal{G}^{\nu}_{\mu} = 2\bigg{[}\mathcal{W}_{f}(\mathcal{F}_{\mu\lambda}\mathcal{F}^{\nu\lambda}) - \delta^{\nu}_{\mu}\mathcal{W}\bigg{]}\\
&\nabla_{\mu} (\mathcal{W}_{f}\mathcal{F}^{\beta\mu}) =0
\end{align}
with $\mathcal{W}_{f} = \frac{\partial\mathcal{W}}{\partial f}$. For Bardeen spacetime the form of $\mathcal{W}_{f}$ is given by,
\begin{align}
\mathcal{W}(f) = \frac{3\mathcal{M}}{|\tilde{g}|\tilde{g}^2}\bigg{[} \frac{\sqrt{2\tilde{g}^2 f}}{1 + \sqrt{2\tilde{g}^2 f}}\bigg{]}^{5/2}
\end{align}

\par

It is important to note that Kerr black hole can be retrieved in the absence of nonlinear electrodynamics i.e $\tilde{g} = f=0$. We aim to constrain both the parameters $a$ and $\tilde{g}$ in the light of astrophysical observations.
\par
In order to calculate the event horizon corresponding to the above spacetime we put $g^{rr}=\Delta=0$ which gives,
\begin{align}\label{horizon_bardeen}
r^2 + a^2 - 2\mathcal{M}r\bigg{(} \frac{r^2}{r^2 + \tilde{g}^2}\bigg{)}^{3/2}=0
\end{align}
We require real and positive solutions of the above equation which enables us to understand the observationally favored magnetic monopole charge parameter $\tilde{g}$. Henceforth, in this paper we will denote the magnetic monopole charge by $g=\tilde{g}^2$.
In the next section we discuss how to derive the luminosity from the accretion disk in the Bardeen spacetime.

\section{Epicyclic frequencies of test particles in a stationary axisymmetric black hole spacetime}\label{S3}

In this section we study the motion of massive test particles around a rotating black hole which is described by a stationary, axisymmetric spacetime possessing reflection symmetry. The metric around such a black hole is given by,
\begin{align}
ds^2=g_{tt}dt^2 + 2g_{t\phi}dt d\phi + g_{\phi\phi}d\phi ^2 + g_{rr}dr^2 +g_{\theta\theta} d\theta^2~,
\label{3-1}
\end{align}
such that $g_{\alpha\beta}=g_{\alpha\beta}(r,\theta)$ and $g_{\mu\nu}(r,\theta)=g_{\mu\nu}(r,-\theta)$. It is clear from \ref{3-1} that the above spacetime has two Killing vectors namely $\partial_t$ and $\partial_\phi$ such that the specific energy $E$ and the specific angular momentum $L$ of the test particles are conserved. One can show from the invariance of the rest mass of the test particles that,
\begin{align}
\dot{r}^2g_{rr}+\dot{\theta}^2g_{\theta\theta} +E^2 U(r,\theta)=-1
\label{3-2}
\end{align}
where $U(r,\theta)=(g^{tt}-2lg^{t\phi}+l^{2}g^{\phi\phi})$ corresponds to the effective potential due to the central black hole in which the test particles moves and $l=L/E$ is the impact parameter. 
We first consider circular equatorial motion of the test particles such that $E^2 U(r_0,\pi/2)=-1$ where $r_0$ corresponds to the radius of the circular orbit.

The angular frequency corresponding to the circular, equatorial motion of test particles in the above background is obtained by solving the equation,
\begin{align}
\label{3-3}
g_{tt,r}+2\Omega  g_{t\phi,r} +\Omega^{2} g_{\phi\phi,r}=0~.
\end{align}
such that
\begin{align}
\label{3-4}
\Omega=\frac{ -g_{t\phi,r}\pm \sqrt{g_{t\phi,r}^2 - g_{tt,r}g_{\phi\phi,r}}}{ g_{\phi\phi,r}}
\end{align}
where $\pm$ sign corresponds to prograde and retrograde orbits.
We now consider slight perturbation in the motion of the test particle from the circular orbit and the equatorial plane. Let us denote this by,
\begin{align}
r(t)\simeq r_{\rm 0}+\delta r_{0}~e^{i\omega_r t}~;
\qquad
\theta(t)\simeq \frac{\pi}{2}+\delta\theta_{0}~e^{i\omega_\theta t}~.
\label{3-5}
\end{align}
where $\omega_r$ and $\omega_\theta$ correspond to radial and vertical epicyclic frequencies respectively.
Substituting \ref{3-5} in \ref{3-2} and Taylor expanding $U(r,\theta)$ about $r=r_0$ and $\theta=\pi/2$,
\begin{align}
-\delta r_{0}^2 \omega_r^2 (u^t)^2g_{rr}-\delta \theta_{0}^2 \omega_\theta^2 (u^t)^2g_{\theta\theta} + E^2 \bigg[U(r_0,\pi/2) + \frac{1}{2}\frac{\partial^2U}{\partial r^2}\bigg|_{r=r_0,\theta=\pi/2} \delta r_0^2+ \frac{1}{2}\frac{\partial^2U}{\partial \theta^2}\bigg|_{r=r_0,\theta=\pi/2}\delta \theta^2 \bigg] =-1
\label{3-6}
\end{align}
We have noted earlier that $E^2 U(r_0,\pi/2) +1=0$ and since the motion along the radial and the vertical direction are uncoupled in the linear approximation we may equate the coefficients of $\delta r^2$ and $\delta \theta^2$ on both sides of \ref{3-6}.
This yields the following expressions for the radial and the vertical epicyclic frequencies:
\begin{align}
\omega_r^2 =\frac{c^6}{G^2M^2} \frac{(g_{tt}+\Omega g_{t\phi})^2}{2g_{rr}}\frac{\partial^2U}{\partial r^2}\bigg|_{r=r_0,\theta=\pi/2}  \\
\omega_\theta^2 = \frac{c^6}{G^2M^2} \frac{(g_{tt}+\Omega g_{t\phi})^2}{2g_{\theta\theta}}\frac{\partial^2U}{\partial \theta^2}\bigg|_{r=r_0,\theta=\pi/2} 
\label{3-7}
\end{align}
The radial and vertical epicyclic frequencies are multiplied by the factor $(c^6/G^2M^2)$ so that they have dimensions of frequency squared.

We next discuss various theoretical models aimed to explain the observed high frequency quasi-periodic oscillations (HFQPOs) in the power spectrum of black holes. HFQPOs are observed 
in some stellar mass black holes and supermassive black holes with frequencies $\sim$ hundreds of Hz and $\sim $ mHz respectively \cite{2008Natur.455..369G,Torok:2004xs,Aschenbach:2004kj}. The order of magnitude of these QPO frequencies can be directly attributed to the masses of the black hole in question (see \ref{3-7}). Apart from the HFQPOs, some of these black holes also exhibit low frequency QPOs (LFQPOs) in their power spectrum. In \ref{Table1} we present a few BH sources where QPOs have been discovered. Interestingly, it turns out that the HFQPOs in the BH sources appear generally in the ratio of 3:2. The two observed HFQPOs are denoted by $f_{u1}$ and $f_{u2}$ while the low frequency QPO is denoted by $f_L$. In this regard it may be important to note that the RE J1034+396 galaxy exhibits only a single QPO in its power spectrum \cite{2008Natur.455..369G,Middleton:2008fe,Jin:2020hgq,Jin:2020meg}. 

Various theoretical models have been proposed to explain the observed QPOs in the black hole power spectrum \cite{Stella:1997tc,PhysRevLett8217,Stella_1999,Cadez:2008iv,Kostic:2009hp,Germana:2009ce,Kluzniak:2002bb,Abramowicz:2003xy,Rebusco:2004ba,Nowak:1996hg,Torok:2010rk,Torok:2011qy,Kotrlova:2020pqy,1980PASJ...32..377K,Perez:1996ti,Silbergleit:2000ck,Dexter:2013sxa,Rezzolla:2003zy,Rezzolla:2003zx}. 
These models chiefly aim to explain the commensurability of the QPO frequencies. The theoretical HFQPOs obtained from various models are denoted by $f_1$ and $f_2$ while the theoretical low frequency QPO is denoted by $f_3$. The mathematical expressions for each of these QPO frequencies from various QPO models are presented in \ref{Table2} from which it is evident that these QPO frequencies are linear combinations of $\nu_\phi=\omega_\phi/2\pi$, $\nu_r=\omega_r/2\pi$ and $\nu_\theta=\omega_\theta/2\pi$.
It may be important to note that the theoretical QPO frequencies depend solely on the background spacetime and not on the complex physics of the accretion processes. Therefore, QPOs can potentially be a cleaner probe to the background spacetime compared to other available observations, e.g., the iron line or the continuum-fitting methods.
Further, we will consider only those sources in \ref{Table1} (namely, the first five sources) which exhibit the 3:2 ratio HFQPOs. Therefore the data related to RE J1034+396 galaxy will not be used for subsequent analysis.

In the present work we assume the black holes mentioned in \ref{Table1} to be Bardeen BHs 
and aim to extract their magnetic monopole charge from QPO related observations.

\begin{table}[h]
\begin{center}
\begin{tabular}{|c|c|c|c|c|}
\hline
$\rm Source $ & $\rm Mass$ & $ f_{u1} \pm  \Delta f_{u1}$ & $ f_{u2} \pm {\Delta} f_{u2}$ & $ f_L \pm \Delta f_L$\\
& $\rm (\rm in ~M_\odot)$ & $(\rm in~ Hz)$ & $\rm (\rm in ~Hz)$ & $\rm (\rm in~ Hz)$\\
\hline 
$\rm GRO ~J1655-40$ & $\rm 5.4\pm 0.3$ \cite{Beer:2001cg} & $\rm 441  \rm \pm 2 $ \cite{Motta:2013wga} & $\rm 298 \rm \pm 4 $ \cite{Motta:2013wga} & $\rm 17.3 \pm \rm 0.1 $ \cite{Motta:2013wga}\\ 
\hline
$\rm XTE ~J1550-564$ & $\rm 9.1\pm 0.61$ \cite{Orosz:2011ki} & $\rm 276 \rm \pm 3 $ & $\rm 184  \pm 5 $ & $ -$\\
\hline
$\rm GRS ~1915+105$ & $\rm 12.4^{+2.0}_{-1.8}$ \cite{Reid:2014ywa} & $\rm 168  \pm 3 $ & $\rm 113  \pm 5 $ & $\rm - $\\
\hline
$\rm H ~1743+322$ & $\rm 8.0-14.07$ \cite{Pei:2016kka,Bhattacharjee:2019vyy,Petri:2008jc} & $\rm 242 \pm 3 $ & $\rm 166  \pm 5 $ & $\rm - $\\
\hline
$\rm Sgr~A^*$ & $\rm (3.5-4.9)$ & $\rm (1.445 \pm 0.16)$ & $\rm (0.886 \pm 0.04)$ & $ - $\\
 & $\rm ~\times 10^6$ \cite{Ghez:2008ms,Gillessen:2008qv} & $\rm ~\times 10^{-3} $ \cite{Torok:2004xs,Stuchlik:2008fy} & $\rm ~\times 10^{-3} $ \cite{Torok:2004xs,Stuchlik:2008fy} & $ - $\\
 \hline
 $\rm RE J1034+396$ & $\rm (1-4) ~\times 10^6$ & $\rm (2.5-2.8) \rm ~\times 10^{-4}$ & $-$ & $ - $\\
 &  \cite{2008Natur.455..369G,2012MNRAS.420.1825J,Czerny:2016ajj,Chaudhury:2018jzz} &  \cite{2008Natur.455..369G,Middleton:2008fe,Jin:2020hgq,Jin:2020meg} &   & \\
\hline
\end{tabular}
\caption{Black hole sources exhibiting high frequency QPOs (HFQPOs)}
\label{Table1}
\end{center}

\end{table}

\begin{table}[h!]
\begin{center}
\begin{tabular}{|c|c|c|c|}
\hline
$\rm Model $ & $ f_1 $ & $ f_2$ &  $f_3 $ \\
\hline 
$\rm Relativistic ~Precession ~Model ~(kinematic)$ \cite{Stella:1997tc,PhysRevLett8217,Stella_1999} & $\rm \nu_\phi$ & $\rm \nu_\phi-\nu_r $ & $\rm \nu_\phi-\nu_\theta$ \\ 
\hline
$\rm Tidal ~Disruption~Model~(kinematic)$ \cite{Cadez:2008iv,Kostic:2009hp,Germana:2009ce} & $\rm \nu_\phi + \nu_r$ & $\rm \nu_\theta $ & $-$ \\
\hline
$\rm Parametric ~Resonance~Model~(resonance)$ \cite{Kluzniak:2002bb,Abramowicz:2003xy,Rebusco:2004ba} & $\rm \nu_\theta$ & $\rm \nu_r$ & $-$\\ \hline
$\rm Forced ~Resonance~Model~1 ~(resonance)$ \cite{Kluzniak:2002bb} & $\rm \nu_\theta$ & $\rm \nu_\theta-\nu_r$ & $-$\\ \hline
$\rm Forced ~Resonance~Model~2 ~(resonance)$ \cite{Kluzniak:2002bb} & $\rm \nu_\theta+ \nu_r$ & $\rm \nu_\theta$ & $-$\\ \hline
$\rm Keplerian ~Resonance~Model~1 ~(resonance)$ \cite{Nowak:1996hg} & $\rm \nu_\phi$ & $\rm \nu_r$ & $-$\\ \hline
$\rm Keplerian ~Resonance~Model~2 ~(resonance)$ \cite{Nowak:1996hg} & $\rm \nu_\phi$ & $\rm 2\nu_r$ & $-$\\ \hline
$\rm Keplerian ~Resonance~Model~3 ~(resonance)$ \cite{Nowak:1996hg} & $\rm 3 \nu_r$ & $\rm \nu_\phi$ & $-$\\ \hline
$\rm Warped ~Disk~Oscillation~Model~ ~(resonance)$ & $\rm 2\nu_\phi-\nu_r$ & $\rm 2(\nu_\phi-\nu_r)$ & $-$\\ \hline
$\rm Non-axisymmetric ~Disk~Oscillation~Model~1 ~(resonance)$ & $\rm \nu_\theta$ & $\rm \nu_\phi-\nu_r$ & $-$\\ \hline
$\rm Non-axisymmetric ~Disk~Oscillation~Model~2 ~(resonance)$  \cite{Torok:2010rk,Torok:2011qy,Kotrlova:2020pqy} & $\rm 2\nu_\phi-\nu_\theta$ & $\rm \nu_\phi-\nu_r$ & $-$\\ \hline
\end{tabular}
\caption{Theoretical expressions for the HFQPOs and the LFQPO from various QPO models.}
\label{Table2}
\end{center}
\end{table}
\section{Theoretical Models explaining quasi-periodic oscillations in the black hole power spectrum}\label{S4}

In this section we review some of the existing theoretical models aimed to explain the observed HFQPOs in BHs. Since the model dependent QPO frequencies (\ref{Table2}) are linear combinations of $\nu_\phi$, $\nu_r$ and $\nu_\theta$, one can easily check that these are functions of the black hole mass $M$, the radius at which these oscillations are generated $r_{em}$ and the metric parameters (which in this case corresponds to $g$ and $a$). The QPO models existing in the literature can be broadly classified into two categories, namely the kinematic models and the resonant models.
Below we briefly review the basic features of each of these models:

\begin{itemize}
\item {\bf Kinematic models:} In kinematic models the origin of QPOs is attributed to the
local motion of plasma in the accretion disk. The two most widely used kinematic models are the Relativistic Precession Model (henceforth referred to as RPM) \cite{Stella:1997tc,PhysRevLett8217} and the Tidal Disruption Model (TDM) \cite{Cadez:2008iv,Kostic:2009hp,Germana:2009ce}. The Relativistic Precession Model, originally proposed to explain the twin HFQPOs in neutron star sources is also used to address the HFQPOs observed in black holes \cite{Stella_1999}. According to this model, the observed upper and lower HFQPOs (denoted by $f_{\rm up1}$ and $f_{\rm up2}$) corresponds to the orbital angular frequency and the periastron precession frequency such that $f_1=\nu_\phi$ and $f_2=\nu_\phi-\nu_r$. This model can also be used to the explain the observed LFQPOs in BHs such that $f_L$ is associated with the nodal precession frequency, i.e., $f_{3}=\nu_{\phi}-\nu_{\theta}$.

Another widely used kinematic model is the Tidal Disruption Model \cite{Cadez:2008iv,Kostic:2009hp,Germana:2009ce}. According to this model, the observed modulation of the flux in the power spectrum leading to HFQPOs is attributed to the plasma orbiting the BH which may get tidally stretched by the central object. In this model, the upper and lower HFQPOs are given by $f_1=\nu_\phi+\nu_r$ and $f_2=\nu_\phi$ respectively. 
Thus, the theoretical QPO frequencies in the above kinematic models depend on the black hole hairs $g$, $a$, and $M$ along with the radius $r_{\rm em}$ at which these oscillations are generated.

\item {\bf Resonant models: } It is interesting to note that the observed twin-peak HFQPOs ($f_{u1}$ and $f_{u2}$) in BH and NS sources occur in the ratio of 3:2. This may indicate that the QPOs might originate as a result of resonance between various oscillation modes in the accretion disk \cite{2001A&A...374L..19A,Kluzniak:2002bb,2001AcPPB..32.3605K}. In this regard, the resonant models are more suitable to explain the commensurability of the QPO frequencies \cite{Torok:2011qy}.

In the previous section we assumed that the accreting matter mostly resides in the the equatorial plane and gradually inspirals and falls into the central black hole in nearly circular geodesics. Slight perturbations from the circular orbit $r_0$ and the equatorial plane are denoted by $\delta r =r-r_{ 0}$ and $\delta \theta=\theta-\pi/2$ respectively, such that they obey the equations corresponding to two uncoupled simple harmonic oscillators with frequencies $\omega_{r}$ and $\omega_{\theta}$,
\begin{align}
\delta \ddot{ r}+\omega_r^2 \delta r=0~; \qquad \delta \ddot{\theta}+\omega_\theta^2 \delta \theta=0~. 
\label{S4-1}
\end{align} 
where $\omega_{r}=2\pi \nu_r$ and $\omega_{\theta}=2\pi \nu_\theta$ correspond to the radial and vertical epicyclic frequencies discussed in \ref{S3}. In a more general situation one needs to incorporate non-linear effects due to pressure and dissipation in the accreting fluid \cite{2001A&A...374L..19A,Kluzniak:2002bb} which requires adding forcing terms on the right hand side of \ref{S4-1}, 
\begin{align}
\delta \ddot{r}+\omega_r^2 \delta r=\omega_r^{2}F_{r}(\delta r,\delta \theta, \delta\dot{r}, \delta\dot{\theta})~; 
\qquad 
\delta \ddot{\theta}+\omega_\theta^2 \delta \theta=\omega_{\theta}^{2} F_{\theta}(\delta r,\delta \theta, \delta\dot{r}, \delta\dot{\theta})~.
\label{S4-2}
\end{align}
In \ref{S4-2} the forcing terms $F_{r}$ and $F_{\theta}$ depend non-linearly on their arguments, their explicit forms being dependent on the accretion flow model. Determining the correct analytical forms of $F_{r}$ and $F_{\theta}$ is non-trivial since dissipation and pressure effects in accretion physics is not very well understood. However, considering various physical situations different analytical forms for the forces $F_{r}$ and $F_{\theta}$ are used \cite{Abramowicz:2003xy, Horak:2004hm} which gives rise to different resonant QPO models. Below we enlist some of the important resonant QPO models:

\begin{enumerate}
\item {\bf Parametric Resonance Model:} Acccording to this model, the radial epicyclic mode induces the vertical epicyclic mode since random fluctuations in thin disks are expected to have $\delta r \gg \delta \theta$ \cite{Kluzniak:2002bb,2001A&A...374L..19A,Abramowicz:2003xy,2005A&A...436....1T,Rebusco:2004ba}. In such a scenario, \ref{S4-2} assumes the form,
\begin{align}
\delta \ddot{ r}+\omega_r^2 \delta r=0 ~~~~~~\delta \ddot{\theta}+\omega_\theta^2 \delta \theta= -\omega_\theta^2 \delta r \delta \theta
\label{S4-3}
\end{align}
From \ref{S4-3} it can be shown that the equation for $\delta \theta$ is similar to the Matthieu equation \cite{1969mech.book.....L} describing parametric resonance and is excited when \cite{Rebusco:2004ba,1969mech.book.....L}
\begin{align}
\frac{\nu_r}{\nu_\theta}=\frac{2}{n}~, 
\qquad
\rm where ~ n \in ~positive ~integers~.
\label{S4-5}
\end{align} 
Since $\nu_\theta > \nu_r$ \cite{Stuchlik:2008fy} for Bardeen black holes, $n=3$ gives rise to the strongest resonance which in turn naturally explains the observed $3:2$ ratio of the HFQPOs. Thus, according to this model, $f_{1}=\nu_\theta$ and $f_{2}=\nu_r$. This finding has been further verified by analytical calculations \cite{Rebusco:2004ba,Horak:2004hm} and numerical simulations \cite{Abramowicz:2003xy}.

\item {\bf{ Forced Resonance Models:}} In general, accretion flows cannot be always modelled by the thin Keplerian disk \cite{2001A&A...374L..19A,Kluzniak:2002bb,2001AcPPB..32.3605K,2005A&A...436....1T} due to the presence of pressure, viscous or magnetic stresses prevelant in the accretion flow. This leads to non-linear couplings between $\delta r$ and $\delta\theta$ in addition to the aforementioned parametric resonance. 
It has been confirmed from numerical simulations that a resonant forcing of vertical oscillations by radial oscillations can be excited through a pressure coupling \cite{2001A&A...374L..19A,2004ApJ...603L..93L}.    
In the absence of an unambiguous understanding of the physics of the accretion flow, such non-linear couplings between $\delta r$ and $\delta\theta$ are often described by some mathematical ansatz, e.g.,
\begin{align}
\delta \ddot{\theta}+\omega_\theta^2 \delta \theta =-\omega_\theta^2\delta r \delta \theta + \mathcal{F_\theta}(\delta\theta) 
\label{S4-7}
\end{align}
such that $\delta r=A cos(\omega_r t)$ while $\mathcal{F_\theta}$ corresponds to the non-linear terms in $\delta\theta$. 
One can show that \ref{S4-7} has solutions of the form,
\begin{align}
\frac{\nu_\theta}{\nu_r}=\frac{m}{n} \rm~~~~~where ~m ~and~ n ~are~ natural~ numbers
\label{S4-8}
\end{align}
We have noted earlier that $m:n=3:2$ is associated with parametric resonance. The presence of non-linear couplings give rise to forced resonances, e.g. $m:n=3:1$ and $m:n=2:1$ which allows resonance between combinations of frequencies, e.g. $\nu_\theta-\nu_r$, $\nu_\theta+\nu_r$.

For $3:1$ forced resonance model (denoted by Forced Resonance Model 1 or FRM1),
the upper HFQPO is given by $f_{1}=\nu_\theta$ while the lower HFQPO is given by $f_{2}=f_-= \nu_\theta-\nu_r$. In case of $2:1$ forced resonance model (denoted by Forced Resonance Model 2 or FRM2) $f_{1}=f_+= \nu_\theta+\nu_r$ while $f_{2}=\nu_\theta$.

\item {\bf{ Keplerian Resonance Models:}} In Keplerian Resonance Models one considers non-linear resonances between the the orbital angular motion and the radial epicyclic modes \cite{2005A&A...436....1T,2001PASJ...53L..37K,2001A&A...374L..19A, Nowak:1996hg}.
Two possible physical scenarios where such resonances might occur corresponds to: (a)
coupling between the radial epicyclic frequencies of a pair of spatially separated coherent vortices with opposite vorticities with their spatially varying orbital angular frequencies \cite{2005A&A...436....1T,2010tbha.book.....A} and/or (b) trapping of non-axisymmetric g-mode oscillations \cite{2001PASJ...53L..37K} in the inner parts of relativistic thin accretion disks. It was however shown \cite{Li:2002yi,2003PASJ...55..257K} that the presence of corotation resonance dampens the g-mode oscillations such that Keplerian resonance models may not be very promising in explaining the HFQPOs. Within the purview of the aforementioned scenarios, the Keplerian Resonance may occur between (a) $f_1=\nu_\phi$ and $f_2=\nu_r$ (denoted by Keplerian Resonance Model 1 or KRM1), (b)  $f_1=\nu_\phi$ and $f_2=2\nu_r$ (denoted by Keplerian Resonance Model 2 or KRM2) and (c) $f_1=3\nu_r$  and $f_2=\nu_\phi$ (which we denoted by Keplerian Resonance Model 3 or KRM3).

\item {{\bf Warped Disk Oscillation Model:}} This model assumes a somewhat unusual disc geometry \cite{Torok:2011qy,Yagi:2016jml} and aims to explain the observed HFQPOs by considering non-linear resonances between the relativistic disk deformed by a warp with various disk oscillation modes \cite{2001PASJ...53....1K,2004PASJ...56..559K,2004PASJ...56..905K,2005PASJ...57..699K,2008PASJ...60..111K}. 
Such resonances include horizontal resonances inducing g-mode and p-mode oscillations as well as vertical resonances which can induce only the g-mode oscillations \cite{2004PASJ...56..559K}. The origin of such resonances can be ascribed to the fact that the radial epicyclic frequency does not exhibit a monotonic variation with the radial distance $r$ \cite{2004PASJ...56..559K}.

\item {{\bf Non-axisymmetric Disk-Oscillation Model:}} Non-axisymmetric Disk-Oscillation models are essentially variants of the Relativistic Precession model \cite{Torok:2011qy}
considering different combinations of non-axisymmetric disc oscillation modes as the origin of the HFQPOs \cite{2004ApJ...617L..45B,2005AN....326..849B,2005ragt.meet...39B,Torok:2011qy,Kotrlova:2020pqy}. These models include non-geodesic effects in the accretion physics originating from the pressure forces by modelling the accretion flow in terms of a slightly non-slender pressure-supported perfect fluid torus \cite{Torok:2015tpu,Sramkova:2015bha,Kotrlova:2020pqy}. 

There exist two main variants of non-axisymmetric disk-oscillation models: the first (denoted by NADO1 or Vertical Precession Resonance Model \cite{2004ApJ...617L..45B,2005AN....326..849B,2005ragt.meet...39B}) considers resonance between the vertical epicyclic frequency ($f_1=\nu_\theta$) with $m=-1$ non-axisymmetric radial epicyclic frequency ($f_2=\nu_\phi-\nu_r$) while the second (denoted by NADO2) assumes coupling between the $m=-2$ non-axisymmetric vertical epicyclic frequency ($f_1=2\nu_\phi-\nu_\theta$) \cite{Torok:2010rk,Torok:2011qy,Kotrlova:2020pqy} with the $m=-1$ non-axisymmetric radial epicyclic frequency ($f_2=\nu_\phi-\nu_r$). Here $m$ refers to the azimuthal wave number of the non-axisymmetric perturbation.

It is interesting to note that using RPM in Kerr geometry the spin of the microquasar GRO J1655-40 does not agree with the predictions of Continuum Fitting/Fe-line method \cite{Motta:2013wga}. However, if NADO1 or Vertical Precession Resonance Model is invoked 
then the spin of GRO J1655-40 is in agreement with the results obtained from the Continuum Fitting method. However, the physical mechanisms inducing couplings between an axisymmetric and a non-axisymmetric mode or between the pairs non-axisymmetric modes are yet not very well-understood \cite{Horak:2008zg,Torok:2011qy}.

\end{enumerate}

It is important to mention that the theoretical models which are considered in the present analysis are mainly aimed to explain the observed high frequency QPOs. These models assume that the background metric should be stationary, axisymmetric and should possess reflection symmetry. Therefore, although these models were originally proposed for the Kerr spacetime, these should hold good for the
Bardeen spacetime as well.
Further, in the present analysis we have considered only those QPO models where the resonances invoked in the theoretical (model-based) QPO frequencies $f_1$ and $f_2$ (and $f_3$ if included within the purview of the chosen model) occur at the same radial distance $r_{em}$ \cite{2016A&A...586A.130S,Yagi:2016jml,Kotrlova:2020pqy} from the black hole. Such an assumption holds good for the kinematic and resonant models discussed above. Apart from these there also exist another class of models, the so called diskoseismic models, where the oscillatory modes giving rise to the twin HFQPOs are generated at different radii from the accretion disk \cite{1980PASJ...32..377K,Perez:1996ti,Silbergleit:2000ck}. We shall not consider such models in this work as they cannot suitably explain the observed 3:2 ratio of HFQPOs 
\cite{Tsang:2008fz,Fu:2008iw,Fu:2010tf}.

In the next section we compare the theoretical QPO frequencies obtained from each of the models presented in \ref{Table2} with the available QPO observations in black holes, namely, the first five sources in \ref{Table1}. These frequencies depend on the mass of the black hole $M$, the emission radius $r_{em}$ and the metric parameters $g$ and $a$. 
Here, we shall not use the QPO related observations to determine the black hole mass, rather we will consider the mass of these BHs estimated earlier from other independent observations, e.g., optical/NIR photometry (see \ref{Table1}). 
The errors associated with the observed QPO frequencies are also reported in \ref{Table1}. We compute the variation of $\chi^2$ with the monopole charge $g$ corresponding to each of the QPO models. This allows us to discern the most favored magnetic monopole charge within the domain of the QPO models considered.

\end{itemize}

\section{Estimating the magnetic monopole charge from the observed QPOs}\label{S5}

In this section we compare the model dependent QPO frequencies enlisted in \ref{Table2} with the observed QPO frequencies in the BH sources. We perform a joint-$\chi^2$ analysis  which enables us to decode the observationally favored values of spin $a$, emission radius $r_{em}$ and the monopole charge parameter $g$ corresponding to each of the BH sources. We emphasize once again that we do not constrain the BH mass from the present analysis but instead use the predetermined masses of these sources from optical/NIR photometry (see \ref{Table1}). \\

We define the $\chi^2$ function in the following way:
\begin{align}
\label{S5-1}
\chi ^2 (g)=\sum_{j} \frac{\lbrace f_{\textrm{u1}{,j}}-f_1(g,a_{\rm min},M_{\rm min},r_{\rm min}) \rbrace ^2}{\sigma_{f_{\rm u1},j}^2}  
+ \sum_{j} \frac{\lbrace f_{\textrm{u2}{,j}}-f_2(g,a_{\rm min},M_{\rm min},r_{\rm min}) \rbrace ^2}{\sigma_{f_{\rm u2}, j}^2}~,
\end{align}
where $f_{\textrm{u1},j}$ and $f_{\textrm{u2},j}$ are respectively the observed upper and lower HFQPOs for the $j^{\rm th}$ source while $f_1$ and $f_2$ are the theoretical model dependent QPO frequencies. The errors in the observed HFQPOs are denoted by 
$\sigma_{f_{\rm u1}, j}$ and $\sigma_{f_{\rm u2}, j}$ also mentioned in \ref{Table1} and $j$ runs from 1 to 5. We note from \ref{Table2} that the mathematical expressions for the theoretical QPO frequencies $f_1$ and $f_2$ depend on the orbital frequency and the epicyclic frequencies which are functions of the metric parameters $g$, $a$, $M$ and the emission radius $r_{\rm em}$. We however wish to determine the most favored value of $g$ from the QPO observations and hence minimize $\chi^{2}$ only with respect to $g$. This requires us to divide the total number of parameters into two classes \cite{1976ApJ...210..642A}, namely, (a) the ``interesting parameters" (in our case $g$), which are obtained from $\chi^2$ minimization and (b) the ``uninteresting parameters" (here $a$, $M$ and $r_{\rm em}$), which are also obtained from $\chi^2$ minimization for various fixed values of the ``interesting parameters".

\begin{figure}[t!]
\centering
\includegraphics[scale=0.58]{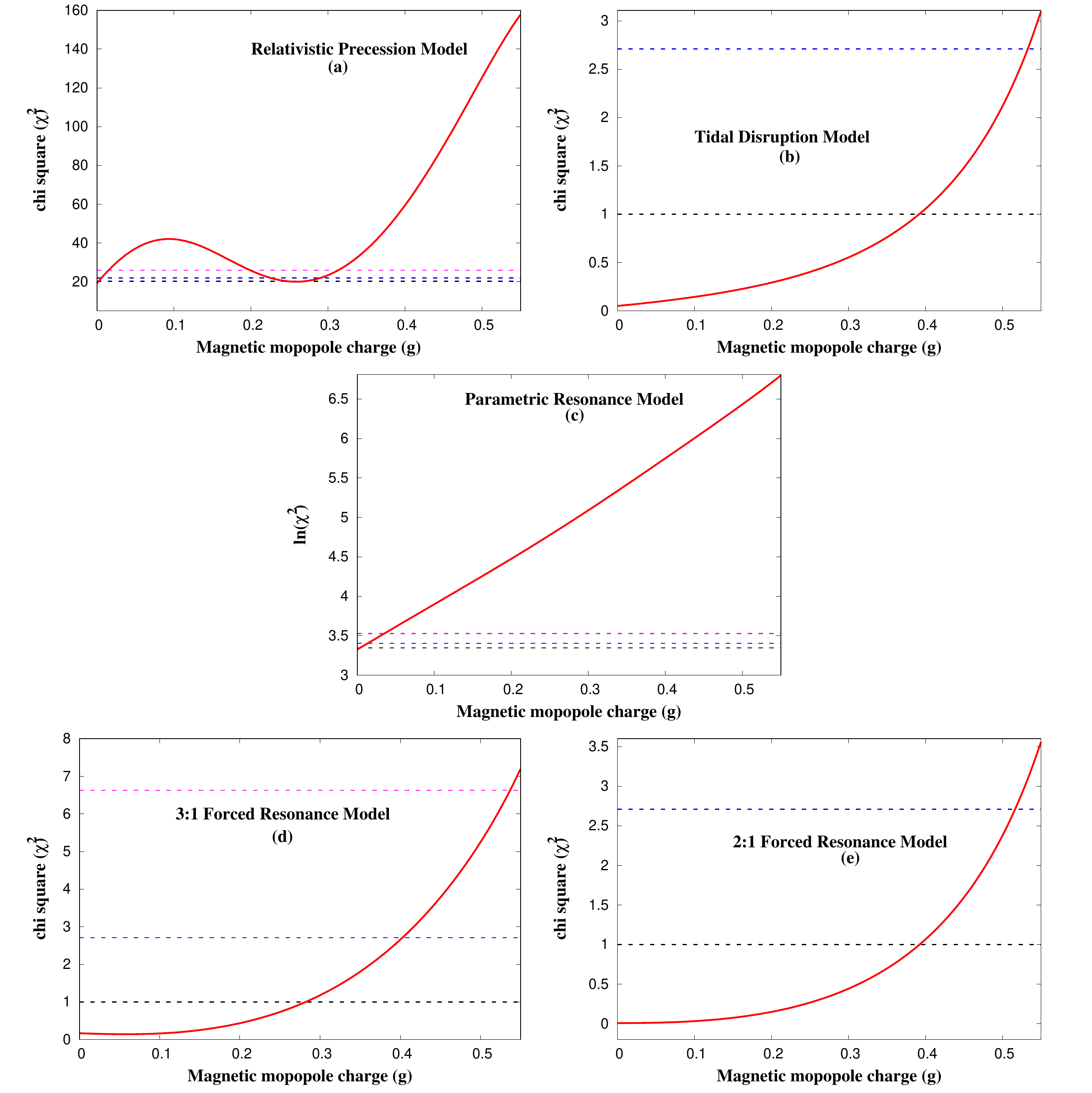}
\caption{In this figure we plot the variation of $\chi^2$ with $g$ assuming the following QPO models: (a) Relativistic Precession Model (RPM), (b) Tidal Disruption Model (TDM), (c) Parametric Resonance Model (PRM), (d) 3:1 Forced Resonance Model (FRM1) and (e) 2:1 Forced Resonance Model (FRM2). $\chi^2$ minimizes for $g=0$ for all the above models indicating that the Kerr scenario in \gr\ is more favored compared to Bardeen rotating black holes.}
\label{Fig_07}
\end{figure}

\begin{figure}[t!]
\centering
\includegraphics[scale=0.58]{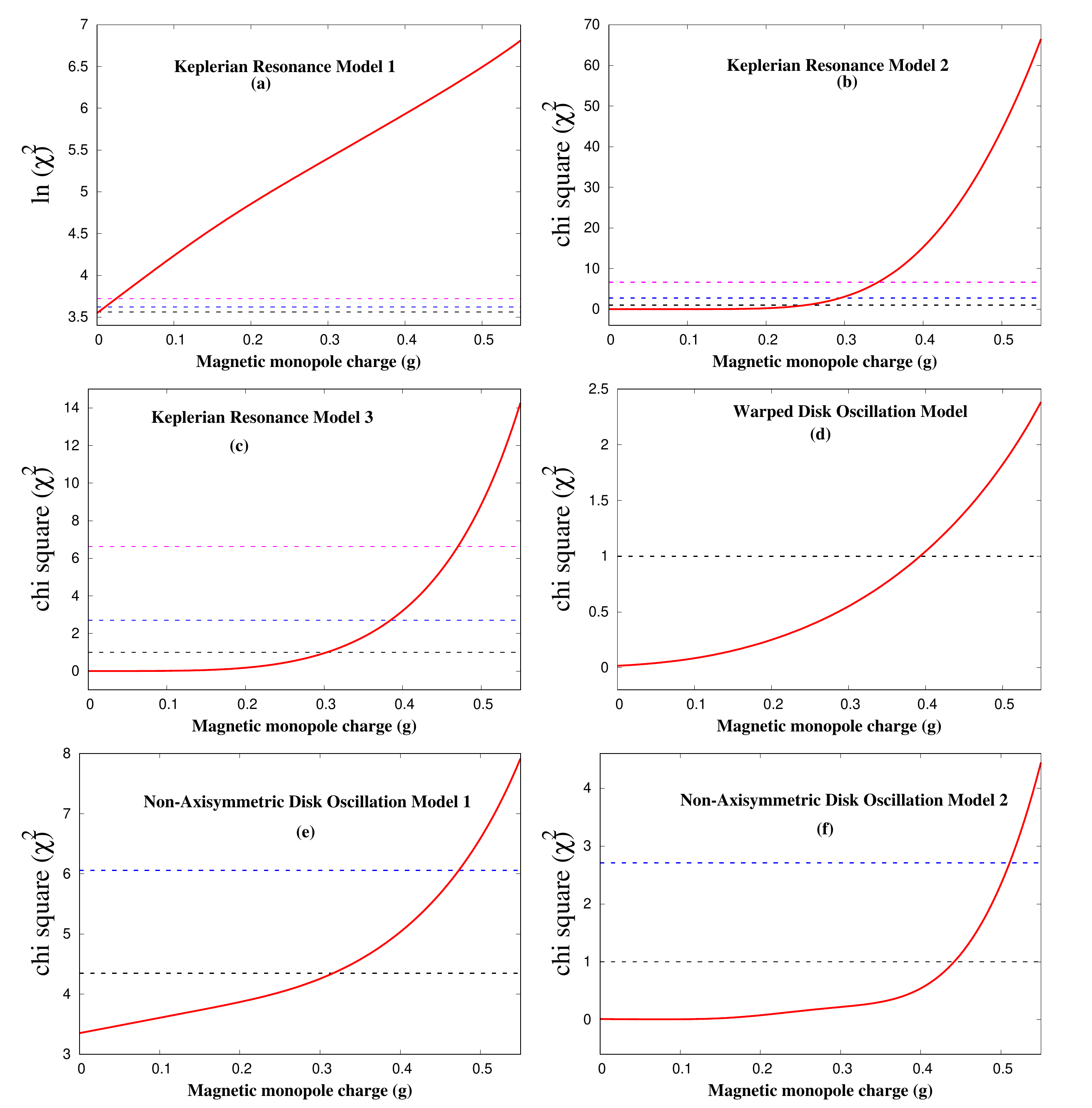}
\caption{The figure above depicts $\chi^{2}$ versus $g$ assuming the following models: (a) Keplerian Resonance Model 1, (b) Keplerian Resonance Model 2, (c) Keplerian Resonance Model 3, (d) Warped Disc Oscillation Model, (e) Non-axisymmetric Disc Oscillation Model 1 and (f) Non-axisymmetric Disc Oscillation Model. Here also $\chi^{2}$ minimizes for $g=0$ consistent with our earlier findings with the previous models. 
}
\label{Fig_08}
\end{figure}

In order to extract the observationally favored value of $g$, we vary the mass of a chosen source within the error bar mentioned in \ref{Table1} (i.e., $(M_{BH}-\Delta M_{BH}) \leq M_{BH} \leq (M_{BH} + \Delta M_{BH})$), the emission radius between $r_{ms}(g,a)\leq r_{em} \leq r_{ms}(g,a) + 20 r_{g}$ and its spin between the maximally allowed prograde and retrograde values for the chosen $g$ and calculate $\chi^2$ as mentioned in \ref{S5-1} considering a given model from \ref{Table2}. The values of mass, $r_{\rm em}$ and spin for which $\chi^2$ minimizes (denoted by $\chi^2_m$) for the chosen source are represented by $M_{\rm min}$, $r_{\rm min}$ and $a_{\rm min}$. We next compute $M_{\rm min}$, $r_{\rm min}$ and $a_{\rm min}$ and hence $\chi^2_m$ for the remaining sources in \ref{Table1} for the same QPO model, keeping $g$ fixed. We repeat the above procedure for other values of $g$ such that $0\leq g \leq 0.55$. This enables us to obtain the variation of $\chi^2_m$ with $g$ for the chosen QPO model. The value of $g$ where $\chi^2_m$ minimizes gives us the observationally favored magnitude of $g$ within the domain of the chosen QPO model. The corresponding value of $\chi^2_m$ is denoted by $\chi^2_{min}$.
The confidence intervals (i.e., $\Delta \chi^{2}$ from $\chi^{2}_{\rm min}$) associated with 68\%, 90\% and 99\% confidence levels are equal to 1, 2.71 and 6.63 \cite{1976ApJ...210..642A}. The variation of $\chi^2$ with $g$ for each of the QPO models is shown in \ref{Fig_07} and \ref{Fig_08}. 
Certain QPO models like RPM can also address the low-frequency QPO observed in GRO J1655-40, in which case the form of $\chi^2$ is given by,
\begin{align}
\label{S5-2}
\chi ^2 (g)&=\sum_{j} \frac{\lbrace f_{\textrm{u1}{,j}}-f_1(g,a_{\rm min},M_{\rm min},r_{\rm min}) \rbrace ^2}{\sigma_{f_{\rm u1}, j}^2}  
+\sum_{j}  \frac{\lbrace f_{\textrm{u2}{,j}}-f_2(g,a_{\rm min},M_{\rm min},r_{\rm min}) \rbrace ^2}{\sigma_{f_{\rm u2}, j}^2} 
\nonumber 
\\
&\hskip 4 cm +\frac{\lbrace f_{L,{\rm GRO}}-f_3(g,a_{\rm min},M_{\rm min},r_{\rm min}) \rbrace ^2}{\sigma_{f_{L},\rm GRO}^2}~.
\end{align}
From the figures it is evident that all the QPO models considered here prefer the general relativistic scenario, since the $\chi^2$ minimizes for $g=0$.

As mentioned earlier we do not provide independent estimates of mass for the given sources, therefore, the value of $M_{min}$ corresponding to the most favored $g$ lies somewhere within the error bars mentioned in \ref{Table1}. The observationally favored values of BH mass from each of the QPO models are reported in \ref{Table3}. 
We however provide independent estimates of spin from the present analysis since (i) the earlier measurements of spin are based on Kerr geometry, while in our case we are considering Bardeen spacetime and (ii) there exist huge disparity in the spin estimates for the same source (e.g., GRO J1655-40 \cite{Motta:2013wga}) when different methods of spin measurements are used assuming \gr. In \ref{Table4} we report the observationally favored values of spin from the present analysis corresponding to $g=0$.



\begin{table}[t!]
\begin{center}
\hspace*{-2cm}
\begin{tabular}{|p{1.8cm}|p{2.8cm}|p{2.8cm}|p{3cm}|p{3cm}| p{3.5cm}| }
\hline

$\rm Comparison$ & $\rm GRO ~J1655-40  $ & $\rm XTE ~J1550-564 $ & $\rm GRS ~1915+105 $ & $\rm H ~1743+322 $ & $\rm Sgr~A^*$\\
$\rm of ~mass $ & $\rm   $ & $\rm  $ & $\rm  $ & $\rm  $ & $\rm $\\
$\rm estimates$ & $\rm   $ & $\rm  $ & $\rm  $ & $\rm  $ & $\rm $\\
$\rm (\rm in ~M_\odot) $ & $\rm   $ & $\rm  $ & $\rm  $ & $\rm  $ & $\rm $\\
\hline 
$\rm $ & $\rm $  & $\rm $ & $\rm  $ & $ \rm $ & $\rm $\\
$\rm Previous $ & $\rm 5.4\pm 0.3$ \cite{Beer:2001cg} & $ \rm 9.1\pm 0.61$ \cite{Orosz:2011ki}  & $ \rm 12.4^{+2.0}_{-1.8} $ \cite{Reid:2014ywa} & $ \rm 8.0-14.07$   & $ \rm (3.5-4.9) \rm \times 10^{-3}$\\ 
$\rm constraints$ & $ $ & $ $ & $ $ & $ $ \cite{Pei:2016kka,Bhattacharjee:2019vyy,Petri:2008jc}  & $  $ \cite{Ghez:2008ms,Gillessen:2008qv} \\

\hline
$\rm $ & $\rm $  & $\rm $ & $\rm  $ & $ \rm $ & $\rm $\\
$\rm RPM$ & $\rm 5.13~(g\sim 0)$  & $\rm 9.3~ (g\sim 0)$ & $\rm 13.99~ (g\sim 0)$ & $ \rm 12.06~(g\sim 0)$ & $\rm 4.0\times 10^6~(g\sim 0)$\\

\hline
$\rm $ & $\rm $  & $\rm $ & $\rm  $ & $ \rm $ & $\rm $\\
$\rm TDM$ & $\rm 5.33~(g\sim0)$  & $\rm 9.13~ (g\sim0)$ & $\rm 12.12 ~(g\sim0)$ & $ \rm 10.73~(g\sim0)$ & $\rm 3.5\times 10^6~(g\sim0)$\\
\hline 
$\rm $ & $ $  & $ $ & $\rm  $ & $ \rm $ & $\rm $\\
$\rm PRM$ & $5.17 \rm ~(g\sim0)$  & $\rm 9.63 ~(g\sim 0)$ & $\rm 13.75 ~(g\sim 0)$ & $ \rm 9.39~(g\sim 0)$ & $\rm 3.5\times 10^6~ (g\sim 0)$\\
\hline
$\rm $ & $\rm $  & $\rm $ & $\rm  $ & $ \rm $ & $\rm $\\
$\rm FRM1$ & $\rm 5.2~(g\sim 0)$  & $\rm  8.51~ (g\sim 0)$ & $\rm 13.2~ (g\sim 0)$ & $ \rm 12.57 ~(g\sim 0)$ & $\rm 3.5 \times 10^6 ~(g\sim 0)$\\
 \hline
 $\rm $ & $\rm $  & $\rm $ & $\rm  $ & $ \rm $ & $\rm $\\
 $\rm FRM2$ & $\rm 5.33 ~(g\sim 0)$  & $\rm  9.27~(g\sim 0)$ & $\rm  11.42~(g\sim 0)$ & $ \rm 9.28~(g\sim 0)$ & $\rm 3.6 \times 10^6~(g\sim 0)$\\
 \hline
$\rm $ & $\rm $  & $\rm $ & $\rm  $ & $ \rm $ & $\rm $\\
$\rm KRM1$ & $\rm 5.17(g\sim 0)$  & $\rm 8.49 (g\sim 0)$ & $\rm 12.13 (g\sim 0)$ & $ \rm 8.0(g\sim 0)$ & $\rm 3.5\times 10^6(g\sim 0)$\\
\hline 
$\rm $ & $\rm $  & $\rm $ & $\rm  $ & $ \rm $ & $\rm $\\
$\rm KRM2$ & $\rm 5.36 ~(g\sim 0)$  & $\rm 8.94 ~(g\sim 0)$ & $\rm 14.1 ~ (g\sim 0)$ & $ \rm 13.08 ~(g\sim 0)$ & $\rm 4.8 \times 10^6~(g\sim 0)$\\
\hline 
$\rm $ & $\rm $  & $\rm $ & $\rm  $ & $ \rm $ & $\rm $\\
$\rm KRM3$ & $5.67 \rm ~(g\sim 0)$  & $\rm 9.65 ~(g\sim 0)$ & $\rm 12.71~ (g\sim 0)$ & $ 10.33\rm ~(g\sim 0)$ & $\rm 4.2 \times 10^6~(g\sim 0)$\\
\hline
$\rm $ & $\rm $  & $\rm $ & $\rm  $ & $ \rm $ & $\rm $\\ 
$\rm WDOM$ & $\rm 5.34 ~(g\sim 0)$  & $\rm 9.34 ~(g\sim 0)$ & $\rm 13.86~ (g\sim 0)$ & $ \rm 11.31 ~(g\sim 0)$ & $\rm 3.5 \times 10^6~ (g\sim 0)$\\
\hline 
$\rm $ & $\rm $  & $\rm $ & $\rm  $ & $ \rm $ & $\rm $\\
$\rm NADO1$ & $\rm 5.15~(g\sim 0)$  & $\rm 9.53 ~(g\sim 0)$ & $\rm  12.4~(g\sim 0)$ & $ \rm 11.13 ~(g\sim 0)$ & $\rm 3.5\times 10^6~ (g\sim 0)$\\
\hline
$\rm $ & $\rm $  & $\rm $ & $\rm  $ & $ \rm $ & $\rm $\\
$\rm NADO2$ & $\rm 5.43~ (g\sim 0)$  & $\rm  9.22 ~(g\sim 0)$ & $\rm  14.04 ~(g\sim 0)$ & $ \rm 14.0~ (g\sim 0)$ & $\rm 4 \times 10^6~(g\sim 0)$\\
\hline\hline

\end{tabular}
\caption{In the above table we report the mass estimates of the BH sources considered in \ref{Table1} obtained from $\chi^2$ minimization. These are compared with earlier estimates.
}
\label{Table3}
\end{center}

\end{table}

\begin{table}[t!]
\vskip-1.8cm
\begin{center}
\hspace*{-2cm}
\begin{tabular}{|p{1.8cm} |p{3cm}|p{3.5cm}|p{3cm}|p{2.8cm}| p{3.3cm}| }
\hline

$\rm Comparison$ & $\rm GRO ~J1655-40  $ & $\rm XTE ~J1550-564 $ & $\rm GRS ~1915+105 $ & $\rm H ~1743+322 $ & $\rm Sgr~A^*$\\
$\rm of ~spin $ & $\rm   $ & $\rm  $ & $\rm  $ & $\rm  $ & $\rm $\\
$ \rm estimates $ & $\rm   $ & $\rm  $ & $\rm  $ & $\rm  $ & $\rm $\\
$ \rm  $ & $\rm   $ & $\rm  $ & $\rm  $ & $\rm  $ & $\rm $\\
\hline 
$\rm Previous$ & $\rm a\sim 0.65-0.75 $ \cite{Shafee_2005} & $ \rm -0.11<a<0.71$ \cite{Steiner:2010bt}  & $ a\sim \rm 0.98 $ \cite{McClintock:2006xd} & $ a=\rm 0.2 \pm 0.3 $ \cite{Steiner:2011kd} & $ \rm a\sim 0.92$ \cite{Moscibrodzka:2009gw}\\ 
$\rm constraints$ & $ \rm a\sim 0.94-0.98$ \cite{Miller:2009cw} &  $ $ & $ a\sim \rm 0.7 $ \cite{2006MNRAS.373.1004M} & $ \rm $   & $ \rm a\sim 0.5 $ \cite{Shcherbakov:2010ki} \\
$\rm $ & $ \rm a=0.29\pm 0.003  $ \cite{Motta:2013wga} & $ \rm a= 0.55^{+0.15}_{-0.22}$ \cite{Steiner:2010bt}  & $ \rm a\sim 0.6-0.98 $ \cite{Blum:2009ez} &   $   $   & $ \rm a=0.9959\pm 0.0005$ \cite{2010MmSAI..81..319A}\\ 
$\rm $ & $ $  & $ $ & $ \rm a\sim 0.4-0.98$ \cite{Mills:2021dxs} &  $  $   & $ \rm a \sim 0.1 $ \cite{Fragione:2020khu} \\
\hline
$\rm $ & $\rm $  & $\rm $ & $\rm   $ & $ \rm $ & $\rm $\\ 
$\rm RPM$ & $\rm 0.3~(g\sim 0)$  & $\rm 0.4~(g\sim 0)$ & $\rm 0.3~ (g\sim 0)$ & $ \rm  0.5~(g\sim 0)$ & $\rm 0.97~ (g\sim 0)$\\
\hline
$\rm $ & $\rm $  & $\rm  $ & $\rm  $ & $ \rm$ & $\rm $\\
$\rm TDM$ & $\rm 0.1~(g\sim 0)$  & $\rm 0.23~(g\sim 0)$ & $\rm -0.1~ (g\sim 0)$ & $ \rm  0.2~(g\sim 0)$ & $\rm 0.99~ (g\sim 0)$\\
\hline
$\rm $ & $\rm $  & $\rm  $ & $\rm  $ & $\rm  $ & $\rm  $ \\
$\rm PRM$ & $\rm 0.9~(g\sim 0)$  & $\rm 0.95~(g\sim 0)$ & $\rm 0.9~ (g\sim 0)$ & $ \rm  0.92~(g\sim 0)$ & $\rm 0.99~ (g\sim 0)$\\
\hline
$\rm $ & $\rm $ & $ $ & $\rm $ & $ \rm $   &  $\rm   $\\
$\rm FRM1$ & $\rm 0.3~(g\sim 0)$  & $\rm 0.34~(g\sim 0)$ & $\rm 0.26~ (g\sim 0)$ & $ \rm  0.6~(g\sim 0)$ & $\rm 0.97~ (g\sim 0)$\\
\hline
 $\rm $ & $\rm  $ & $\rm  $ & $ \rm $ & $ $ &  $ $ \\
$\rm FRM2$ & $\rm 0.1~(g\sim 0)$  & $\rm 0.25~(g\sim 0)$ & $\rm -0.2~ (g\sim 0)$ & $ \rm 0.0 ~(g\sim 0)$ & $\rm 0.97~ (g\sim 0)$\\
 \hline
$\rm $ & $\rm  $  & $\rm  $ & $\rm  $ & $\rm  $  &  $ $ \\
$\rm KRM1$ & $\rm 0.99~(g\sim 0)$  & $\rm 0.99~(g\sim 0)$ & $\rm 0.97~ (g\sim 0)$ & $ \rm  0.99~(g\sim 0)$ & $\rm 0.99~ (g\sim 0)$\\
\hline 
$\rm $ & $\rm  $  & $\rm  $ & $\rm  $ & $\rm  $  &  $ $ \\
$\rm KRM2$ & $\rm 0.32~(g\sim 0)$  & $\rm 0.36~(g\sim 0)$ & $\rm 0.32~ (g\sim 0)$ & $ \rm  0.6~(g\sim 0)$ & $\rm 0.97~ (g\sim 0)$\\
\hline 
$\rm $ & $\rm  $  & $\rm $ & $\rm  $ & $\rm  $  &  $ $\\
$\rm KRM3$ & $\rm 0.22~(g\sim 0)$  & $\rm 0.3~(g\sim 0)$ & $\rm 0.0~ (g\sim 0)$ & $ \rm  0.22~(g\sim 0)$ & $\rm 0.99~ (g\sim 0)$\\
\hline 
$\rm $ & $\rm $ & $\rm  $ & $\rm  $ & $\rm  $  &  $ \rm  $\\
$\rm WDOM$ & $\rm 0.1~(g\sim 0)$  & $\rm 0.26~(g\sim 0)$ & $\rm 0.1~ (g\sim 0)$ & $ \rm  0.26~(g\sim 0)$ & $\rm 0.99~ (g\sim 0)$\\
\hline 
$\rm $ & $\rm $  & $\rm  $ & $\rm  $ & $\rm  $  &  $\rm  $ \\
$\rm NADO1$ & $\rm 0.36~(g\sim 0)$  & $\rm 0.6~(g\sim 0)$ & $\rm 0.22~ (g\sim 0)$ & $ \rm  0.6~(g\sim 0)$ & $\rm 0.99~ (g\sim 0)$\\
\hline 
$\rm $ & $\rm $  & $\rm  $ & $\rm  $ & $\rm  $   &  $\rm  $ \\
$\rm NADO2$ & $\rm 0.25~(g\sim0)$  & $\rm 0.31~(g\sim 0)$ & $\rm 0.24~ (g\sim 0)$ & $ \rm 0.5 ~(g\sim 0)$ & $\rm 0.8~ (g\sim 0)$\\
\hline

 \hline
 \end{tabular}
\caption{
In the above table we report the spin estimates of the BH sources considered in \ref{Table1} obtained from $\chi^2$ minimization. The spin measurements obtained from earlier estimates are also reported.
}
\label{Table4}
\end{center}

\end{table}

We note from \ref{Table4} that previous estimates of spin for sources like GRO J1655-40, GRS 1915+105 and Sgr A* assuming Kerr spacetime exhibit a lot of discrepancy, e.g., for the source GRS 1915+105 the Fe-line method gives a spin $a \sim 0.6- 0.98$ \cite{Blum:2009ez} while the Continuum Fitting Method itself yields maximal ($a\sim 0.98$ \cite{McClintock:2006xd}) as well as intermediate spin ($a\sim 0.7$ \cite{2006MNRAS.373.1004M}). When revised mass and inclination for this source is considered the spin of GRS 1915+105 turns out to be $0.4<a<0.98$. Our results assuming KRM1 and PRM are in best agreement with previous studies \cite{McClintock:2006xd,Blum:2009ez,Mills:2021dxs} (evident from Column 4 of \ref{Table4}). 
For the source GRO J1655-40, assuming Fe-line method the spin turns out to be $0.94<a<0.98$ \cite{Miller:2009cw}, from the Continuum-Fitting method the spin of GRO J1655-40 is $0.65<a<0.75$ \cite{Shafee_2005} while from QPO related observations and assuming RPM the spin is given by $a=0.290\pm 0.003$ \cite{Motta:2013wga}. We note that our spin estimates based on FRM1 and KRM2 are consistent with that of previous spin measurements assuming RPM. Further, the spin of the source that we have obtained assuming PRM and KRM1 are consistent with the results assuming the Fe-line method \cite{Miller:2009cw}. The other QPO models however yield spin values which are different from earlier measurements.
When we consider the source XTE J1550-564, the Fe-line method yields $a=0.55^{+0.15}_{-0.22}$ \cite{Steiner:2010bt} while the Continuum Fitting method gives a spin in the range $-0.11<a<0.71$. Our results assuming most of the QPO models are in agreement with previous results, except PRM and KRM1 (see e.g., third column of \ref{Table4}). For the source H1743-322 the upper bound on spin turns out to be $a<0.92$ (with 99.7\% confidence) while $a=0.2\pm 0.3 $ (with 68\% confidence) when the the Continuum-Fitting method is used \cite{Steiner:2011kd}. Once again apart from PRM and KRM1 all the other QPO models yield results consistent with earlier measurements.
The spin of Sgr A* is obtained from its radio spectrum \cite{Reynolds:2013rva,Moscibrodzka:2009gw,Shcherbakov:2010ki}. Due to the difficulties in modelling the radio spectrum of Sgr A*, there exist a lot of discrepancy in its spin estimates e.g. $a\sim 0.9$ \cite{Moscibrodzka:2009gw}, $a\sim 0.5$ \cite{Shcherbakov:2010ki}. By studying the motion of S2 stars near Sgr A* one obtains $a\lesssim 0.1$ \cite{Fragione:2020khu} while from the X-ray light curve of Sgr A* one infers a maximal spin for this object ($ \rm a=0.9959\pm 0.0005$) \cite{2010MmSAI..81..319A}. Our analysis reveals that the source has a high spin from all the QPO models ($a\geq 0.8$ sixth column of \ref{Table4}) which is consistent with \cite{Moscibrodzka:2009gw,2010MmSAI..81..319A}.

In \ref{Fig_07} and \ref{Fig_08} we plot $\chi^{2}$ as a function of the magnetic monopole charge parameter $g$ for the QPO models discussed in the last section. In particular \ref{Fig_07} depicts the variation of $\chi^2$ with $g$ for: (i) Relativistic Precession Model (RPM), (ii) Tidal Disruption Model (TDM), (iii) Parametric Resonance Model (PRM), (iv) 3:1 Forced Resonance Model (FRM1) and (v) 2:1 Forced Resonance Model (FRM2) while \ref{Fig_08} shows the variation of $\chi^2$ with $g$ for (i) Keplerian Resonance Model 1 (KRM1), (ii) Keplerian Resonance Model 2 (KRM2), (iii) Keplerian Resonance Model 3 (KRM3) (iv) Warped Disc Oscillation Model (WDOM) (v) Non-axisymmetric Disc Oscillation Model 1 (NADO1) and (vi) Non-axisymmetric Disc Oscillation Model 2 (NADO2).
The value of $g$ where $\chi^{2}$ minimizes is denoted by $g_{min}$ and is the one favoured by observations. For each of the QPO models we plot the $1-\sigma$, $2-\sigma$ and $3-\sigma$ confidence intervals which are denoted by $\chi^2=\chi^2_{\rm min}+1$, $\chi^2=\chi^2_{\rm min}+2.71$ and $\chi^2=\chi^2_{\rm min}+6.63$, respectively. These are plotted with black, blue and magenta dashed lines in \ref{Fig_07} and \ref{Fig_08}. 

We note from \ref{Fig_07} and \ref{Fig_08} that all the eleven QPO models favor the Kerr scenario in \gr\, i.e. $g=0$. In fact the two most widely used QPO models PRM and RPM can establish very strong constraints on the monopole charge parameter $g$, i.e., these models rule out $g\gtrsim 0.03$ outside $99\%$ confidence interval. Interestingly, for RPM, the $\chi^2$ minimizes for $g\simeq 0$ as well $g\simeq 0.25$. We consider $g\simeq 0$ as the global minima and compute the $\Delta\chi^2$ accordingly. It may be important to note that even if we consider $\chi^2_{min}$ corresponding to $g\simeq 0.25$ the bounds on the monopole charge parameter turns out to be very stringent when RPM is considered and like all the other QPO models very high values of $g$ seem to be disfavored from observations associated with QPOs. Interestingly, our present result is in agreement with our previous findings where we compare the theoretical spectrum from the accretion disk with the optical observations of quasars and note that the Kerr scenario is more favored compared to the Bardeen black holes \cite{Banerjee:2021nza}.

\section{Concluding Remarks}\label{S6}

In this paper we study the signatures of magnetic monopole charge from the observed quasi-periodic oscillations (QPOs) in black holes. Such black holes (known as Bardeen black holes in the literature) are interesting as they can evade the $r=0$ singularity present in general relativistic black holes. Astrophysical observations, e.g., the broadened and skewed iron-line, the continuum spectrum, the black hole shadow and the quasi-periodic oscillations can be used to extract information regarding the nature of the background spacetime. In this regard, it may be important to note that QPOs and the black hole shadow are cleaner probes to the background spacetime as these are not so much dependent on the complex physics associated with the accretion flow. The role of the magnetic monopole charge on the black hole shadow has been investigated in \cite{Kumar:2018ple}. In this paper we aim to investigate the role of the magnetic monopole charge in explaining the observed quasi-periodic oscillations in black holes.  

Various models have been proposed in the literature to explain the observed QPO frequencies which are dependent on the epicyclic motion of test particles and hence on the background metric. We calculate the theoretical model dependent QPO frequencies in the Bardeen background which in turn are compared with the available observations. By performing a $\chi^2$ analysis between the theoretical and the observed QPO frequencies we arrive at the conclusion that the Kerr scenario in \gr\ is more favored than black holes with a monopole charge. Interestingly, this result more or less holds good for all the eleven QPO models considered here. In particular, models like PRM, RPM and KRM1 establish very strong constraints on the magnitude of the magnetic monopole charge such that $g\gtrsim 0.03$ is outside $99\%$ confidence interval. This result is in agreement with our previous findings where we derive the theoretical spectrum from the accretion disk and compare it with the optical observations of quasars and note that the Kerr scenario is more favored compared to the black holes with a magnetic monopole charge. However, the Kerr metric turns out to be black hole solution of several alternative gravity model \cite{Sen:1992ua,Campbell:1992hc,Psaltis:2007cw} and hence our present result does not uniquely favor the general relativistic scenario compared to other modified gravity theories.

Before concluding we mention some of the limitations of our present analysis. First, our results are model dependent, i.e. the physical mechanism giving rise to the QPOs is not very clearly understood such that despite observing HFQPOs for decades there exist no common concensus regarding the prefered choice of the QPO model that best describes the observations. In fact, it may as well happen that some black hole-accretion disk (BH-AD) systems are described by e.g. PRM while there may be other BH-AD systems which are best described by RPM and so on. Second, our observational sample is small since not many black holes exhibit HFQPOs in their power spectrum. A larger observational sample leads to improved statistics and hence can establish much stronger constraints on the monopole charge parameter $g$. Finally, the available data is not very precise and with the launch of the ESA (European Space Agency) X-ray mission LOFT (Large Observatory for X-ray Timing) one can expect an improvement in the precision by an order of magnitude which in turn will enhance the scope to investigate the QPO phenomenology in near future.

\section*{Acknowledgements}

The author acknowledges Sumanta Chakraborty and Soumitra SenGupta for insightful discussion during the course of this work. Research of I.B. is funded by the Start-Up
Research Grant from SERB, DST, Government of India
(Reg. No. SRG/2021/000418).

\bibliography{new-ref,KN-ED,QPO,Brane,IB,Gravity_3_partial,Gravity_1_full,Black_Hole_Shadow,EMDA-Jet}

\bibliographystyle{./utphys1}
\end{document}